\documentclass[aps,prl,twocolumn,floatfix,superscriptaddress]{revtex4-1}
\usepackage{graphicx,txfonts,bm}
\usepackage[colorlinks,citecolor=magenta,linkcolor=blue]{hyperref}

\begin{document}

%% title
%%%%%%%%%%%%%%%%%%%%%%%%%%%%%%%%%%%%%%%%%%%%%%%%%%%%%%%%%%%%%%%%%%%%%%%%%%

\title{Signatures of Hundness in Kagome Metals}

%% author list
%%%%%%%%%%%%%%%%%%%%%%%%%%%%%%%%%%%%%%%%%%%%%%%%%%%%%%%%%%%%%%%%%%%%%%%%%%

\author{Li Huang}
\email{lihuang.dmft@gmail.com}
\affiliation{Science and Technology on Surface Physics and Chemistry Laboratory, P.O. Box 9-35, Jiangyou 621908, China}

\author{Haiyan Lu}
\affiliation{Science and Technology on Surface Physics and Chemistry Laboratory, P.O. Box 9-35, Jiangyou 621908, China}

\date{\today}

%% abstract
%%%%%%%%%%%%%%%%%%%%%%%%%%%%%%%%%%%%%%%%%%%%%%%%%%%%%%%%%%%%%%%%%%%%%%%%%%

\begin{abstract}
By means of the density functional theory in combination with the dynamical mean-field theory, we tried to examine the electronic structure of hexagonal FeGe, in which the Fe atoms form a quasi-2D layered Kagome lattice. We predict that it is a representative Kagome metal characterized by orbital selective Dirac fermions and extremely flat bands. Furthermore, Fe's 3$d$ electrons are strongly correlated. They exhibit quite apparent signatures of electronic correlation induced by Hund's rule coupling, such as sizable differentiation in band renormalization, non-Fermi-liquid behavior, spin-freezing state, and spin-orbital separation. Thus, FeGe can be regarded as an ideal platform to study the interplay of Kagome physics and Hundness.  
\end{abstract}

%% make title
%%%%%%%%%%%%%%%%%%%%%%%%%%%%%%%%%%%%%%%%%%%%%%%%%%%%%%%%%%%%%%%%%%%%%%%%%%

\maketitle

%% introduction
%%%%%%%%%%%%%%%%%%%%%%%%%%%%%%%%%%%%%%%%%%%%%%%%%%%%%%%%%%%%%%%%%%%%%%%%%%

%% P1: Strong electronic correlation. Mottness vs. Hundness.
Correlated electron materials manifest a plethora of mysterious and fascinating properties, including Mott insulator-metal transitions, colossal and giant magnetoresistance effects, unconventional superconductivity, strange metals, quantum spin liquids, and non-Fermi-liquid behaviors, just to name a few~\cite{RevModPhys.73.797,RevModPhys.89.025003,RevModPhys.70.1039,RevModPhys.78.17}. Hence they are objects of extensive experimental and theoretical interest. Recently, some people realize that the physical properties of strongly correlated multi-orbital systems or materials are regulated by not only the strength of effective interactions that act on the valence electrons, but also their type that leads to different mechanisms towards strong correlation~\cite{georges:2013}. If the correlated electron materials are in the proximity of Mott insulating state, electron correlation is usually caused by strong on-site Coulomb repulsion $U$, which would suppress the charge fluctuations and localize the valence electrons. This is the essential idea of Mott physics or Mottness~\cite{RevModPhys.78.17,PHILLIPS20061634}. Even though many correlated electron materials are far away from the Mott insulating state and the Coulomb repulsion $U$ is moderate, strong electron correlation is still observed experimentally~\cite{Miao2019,PhysRevLett.114.236401}. In that case, Hund's rule coupling $J_{\text{H}}$, which favors to align the spins in different orbitals, provides an alternative way to yield electron correlation~\cite{Haule_2009}. This is the so-called Hundness~\cite{STADLER2019365,deng:2019}. Therefore, according to the origin of electron correlation, strongly correlated multi-orbital systems or materials can be roughly classified into three categories: Mottness (such as the cuprates and $3d$ transition metal oxides)~\cite{RevModPhys.78.17,RevModPhys.70.1039}, Hundness (notably the ruthenates, iridates, and iron-based superconductors)~\cite{PhysRevB.86.195141,PhysRevB.91.041110,RevModPhys.87.855,Yin2011,Kostin2018}, and a combination of both. Note that their excitation spectra, transport properties, and (spin, orbital, and charge) susceptibilities should exhibit quite distinct temperature dependences and energy scales that quantify the beginning and the completion of screening at spin and orbital channels~\cite{deng:2019}.

%P2: Kagome materials and Kagome metals.
In this paper we focus on a special class of quantum materials, namely the Kagome metals~\cite{PhysRevB.45.12377}. These metals or conductive materials are made to resemble some kind of Kagome pattern, i.e. a two-dimensional network of trihexagonal tiling, at the atomic scale. It is well known that a simple tight-binding model with nearest-neighbour hopping on a Kagome lattice already exposes some unusual features, such as the coexistence of topologically protected linearly dispersive bands and dispersionless flat bands~\cite{Ye2018}. However, the experimental realization of Kagome metals are rare and their band structures are elusive until now. Only two years ago, the long-sought Kagome metals were firstly discovered in a ferromagnetic $3d$-electron metal Fe$_{3}$Sn$_{2}$, which are characterized by Dirac cones and extremely flat bands near the Fermi level~\cite{Ye2018,PhysRevLett.121.096401,Ye2019,Wang_2020,Yin2018}. Later, similar electronic structures have been identified in magnetic metals Fe$_{3}$GeTe$_{2}$~\cite{Zhangeaao6791,Kim2018}, Co$_{3}$Sn$_{2}$S$_{2}$~\cite{Liu2018,Wang2018,Yin2019,PhysRevLett.124.077403}, FeSn~\cite{zl:2020,Kang2020}, and nonmagnetic CoSn~\cite{ws:2020,rc:2020}. Due to unique combination of geometry frustration, spin-orbit coupling, and competing magnetic interactions, Kagome metals are expected to host exotic quantum magnetic states such as frustration-driven quantum spin liquid states~\cite{PhysRevB.45.12377,RevModPhys.89.025003}, novel topological phases such as magnetic Weyl fermions~\cite{Kuroda2017,Wang2018}, and abnormal transport properties such as giant anomalous Hall effects~\cite{Liu2018,Wang2018,Kim2018}. 

%P3: The main results of the present work.
In the available Kagome metals, the Kagome lattices consist of Mn, Fe or Co atoms. Notice that their 3$d$ orbitals are not fully occupied and the effective interactions among their valence electrons are not trivial. Undoubtedly, these Kagome metals belong to some sort of correlated electron materials as well. However, in the previous literatures that concerning with the electronic structures of Kagome metals, the electron correlation among Mn-, Fe- or Co-3$d$ valence electrons is usually ignored~\cite{PhysRevLett.124.077403,Kang2020}. It is still unclear how the 3$d$ electron correlation would modulate the Kagome-derived bands. Besides, the origin of electron correlation in Kagome metals remains enigmatic. Keeping these obstacles in mind, we studied the electronic structure of hexagonal FeGe, which is isostructural with and adjacent to FeSn~\cite{Kang2020}, by using a first-principles many-body approach. Our results reveal that it is a typical Kagome metal with strong electron correlation induced by Hundness.    

\begin{figure*}[ht]
\centering
\includegraphics[width=\textwidth]{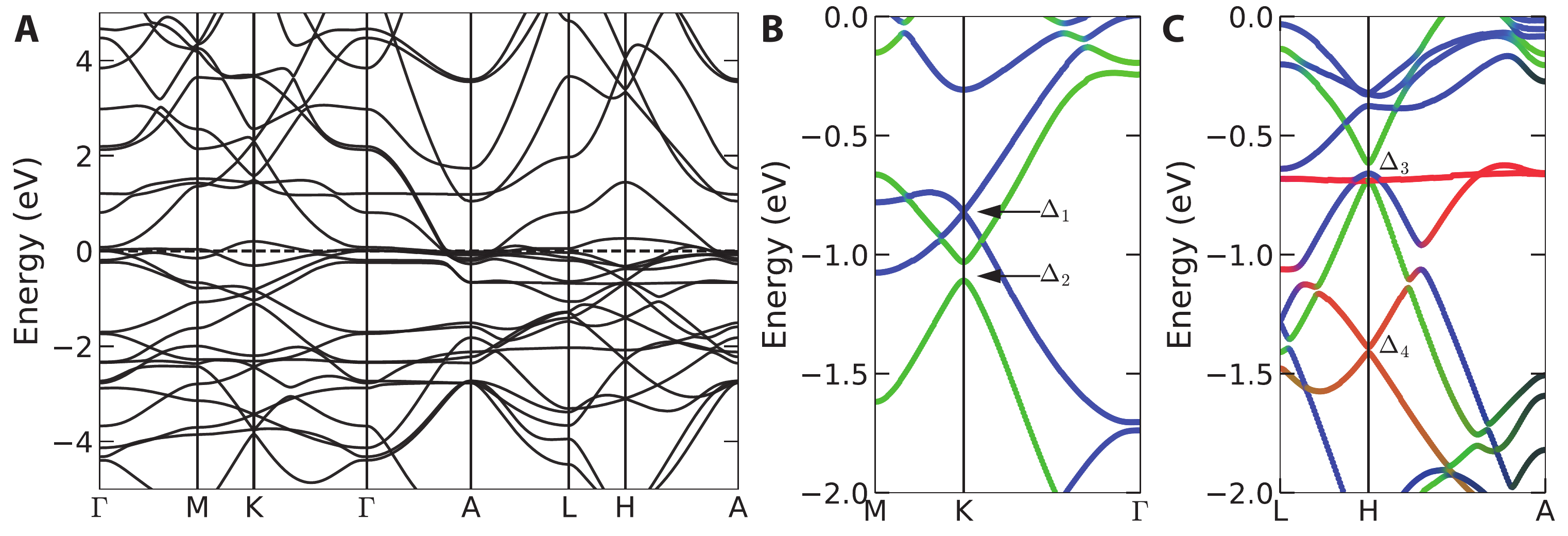}
\caption{(Color online). Band structure of hexagonal FeGe obtained by DFT + SOC calculations. (a) Full band structure along some selected high-symmetry lines in the Brillouin zone. (b)-(c) Enlarged views for the bands near the Fermi level. Here, colors are used to distinguish different orbital characteristics (Red: $d_{z^2}$; Green: $d_{x^2-y^2} + d_{xy}$; Blue: $d_{xz} + d_{yz}$), $\Delta_1 \sim \Delta_4$ mark the Dirac band gaps opened by SOC. \label{fig:band}}
\end{figure*}

%P4: Introduce FeGe.
The hexagonal transition metal stannides $T$Sn (where $T$ = Mn, Fe, or Co; space group: $P_{6}/mmm$) have been proposed to realize the metallic Kagome lattices~\cite{Ye2018,GIEFERS2006132}. The stacking sequences of these intermetallic compounds are $A-S-A$, where $A$ denotes Kagome layers and $S$ means spacing layers. The Kagome networks consist of $T$ atoms and the centers of the hexagons are populated by Sn atoms. The spacing layers consist of Sn atoms in a hexagonal arrangement. If Sn atoms in FeSn are substituted by Ge atoms, we get FeGe, which has a smaller lattice constant. We studied the electronic structure of FeGe using the density functional theory (DFT). In order to treat the electronic correlation in Fe-3$d$ orbitals, the dynamical mean-field theory (DMFT) was employed~\cite{RevModPhys.78.865,RevModPhys.68.13,theory}. 

%% results and discussion
%%%%%%%%%%%%%%%%%%%%%%%%%%%%%%%%%%%%%%%%%%%%%%%%%%%%%%%%%%%%%%%%%%%%%%%%%%

%P5: Orbital selective Dirac fermions.
\emph{Kagome-derived Dirac fermions.} Fig.~\ref{fig:band}(a) shows the \emph{ab initio} band structure of hexagonal FeGe. The following characteristics are quite prominent: (i) Low-dispersion or even non-dispersing flat bands run through the whole Brillouin zone. Especially, a flat band almost pins at the Fermi level along the $\Gamma-M$ and $\Gamma-A-L$ directions. (ii) Linearly dispersive Dirac bands emerge. They form multiple Dirac cones at the $K$ and $H$ points. Both are the defining features of an ideal Kagome metal~\cite{Kang2020}. 

%P6: Orbital selective Dirac fermions.
If the spin-orbit coupling (SOC) is considered explicitly in the calculations, small gaps will be opened at the Dirac points. As a consequence, the linearly dispersive bands are supplemented by a mass term and massive Dirac fermions are generated~\cite{Ye2018}. For example, in Fig.~\ref{fig:band}(b)-(c), four Dirac gaps ($\Delta_{1} \sim \Delta_{4}$) are clearly labelled. Here, we would like to emphasize that the generation of massive Dirac fermions is orbital selective. Under the hexagonal crystal field, Fe's five 3$d$ orbitals are split into three groups, namely $d_{z^2}$, $d_{x^2-y^2} + d_{xy}$, and $d_{xz} + d_{yz}$. The $d_{z^2}$ and $d_{xz} + d_{yz}$ orbitals are out-of-plane, while the $d_{x^2-y^2} + d_{xy}$ orbitals are in-plane. Since the SOC strengths of the in-plane orbitals ($\lambda_{x^2 - y^2}$ and $\lambda_{xy}$) are much stronger than those of the out-of-plane orbitals ($\lambda_{xz}$, $\lambda_{yz}$, and $\lambda_{z^2}$), the sizes of the Dirac gaps opened by these orbitals are quite different. The Dirac gaps with $d_{z^2}$ orbital character ($\Delta_4$ = 58.8~meV) and with $d_{xz} + d_{yz}$ orbital characters ($\Delta_1$ = 10.5~meV) are much smaller than those with the $d_{x^2-y^2} + d_{xy}$ orbital characters ($\Delta_2$ = 78.8~meV and $\Delta_3$ = 72.8~meV). These results suggest that the orbital differentiation in SOC strengths is remarkable, which will finally lead to orbital selective Dirac fermions and topological excitations. We are aware that similar results have been observed in nonmagnetic Kagome metal CoSn~\cite{ws:2020}.

\begin{figure*}[ht]
\centering
\includegraphics[width=\textwidth]{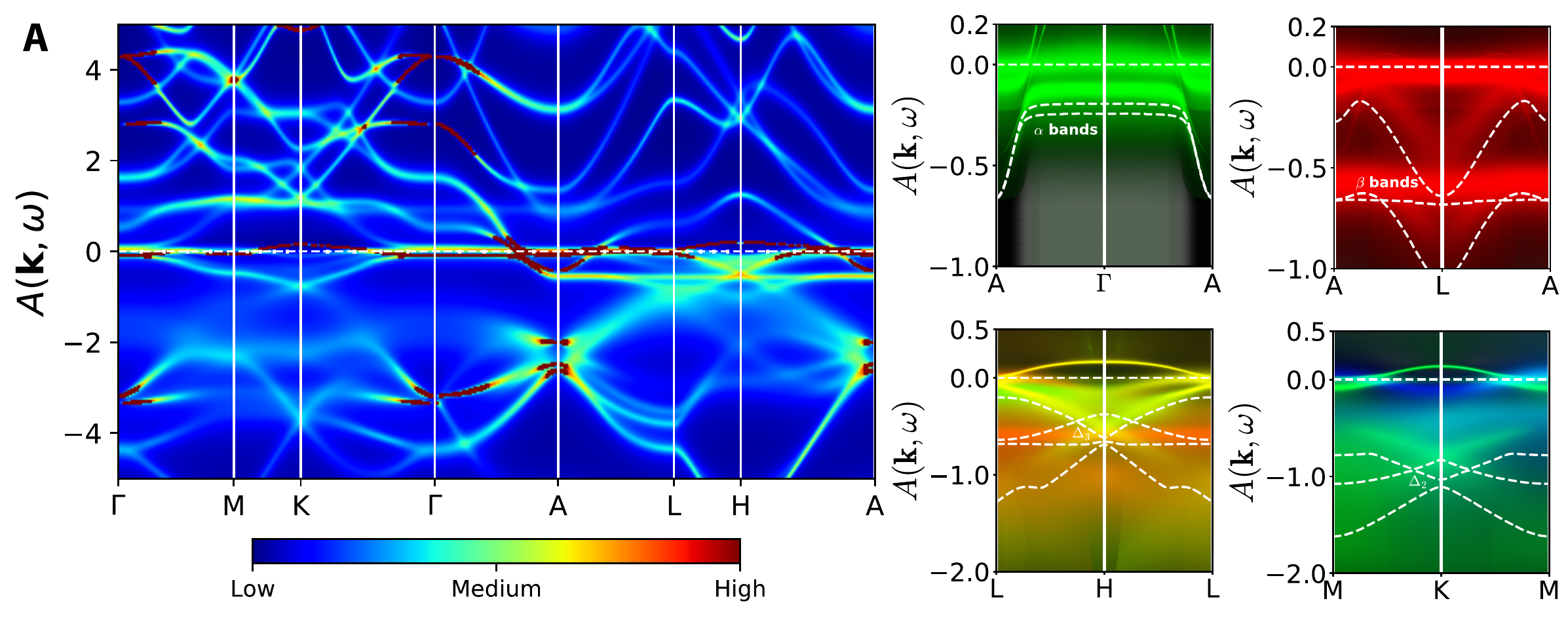}
\caption{(Color online). Momentum-resolved spectral functions of hexagonal FeGe obtained by DFT + DMFT calculations. (a) Full $A(\mathbf{k},\omega)$ along some selected high-symmetry directions in the Brillouin zone. (b)-(e) Orbital-resolved $A(\mathbf{k},\omega)$. Here, colors are used to highlight different orbital characteristics (Red: $d_{z^2}$; Green: $d_{x^2-y^2} + d_{xy}$; Blue: $d_{xz} + d_{yz}$). The thick dashed lines denote the DFT + SOC bands and $E_{F}$. \label{fig:akw}}
\end{figure*}

%P7: Kagome flat bands.
\emph{Kagome-derived flat bands.} Multiple flat bands over a wide range of momentum are intrinsic features of Kagome metals~\cite{PhysRevB.45.12377,Ye2018,PhysRevLett.121.096401,Kang2020,rc:2020,ws:2020}. These dispersionless bands imply highly degenerate manifold state of electrons, ultra singular density of states, extremely heavy localized electrons~\cite{PhysRevLett.121.096401}. They may provide unique opportunities for the emergence of intriguing quantum states of matter. However, what is the origin of these flat bands? According to previous calculations and modelings, the band flatness is tightly connected to the local destructive interferences of Bloch wave functions within the Kagome lattices, instead of electron correlation~\cite{PhysRevLett.121.096401}. In spite of this, it is useful to clarify the influence of strong electron correlation on the electronic structure of FeGe.      

%P8: Kagome flat bands. 
Fig.~\ref{fig:akw}(a) shows the momentum-resolved spectral functions $A(\mathbf{k},\omega)$ of hexagonal FeGe obtained by the DFT + DMFT method~\cite{RevModPhys.78.865,RevModPhys.68.13,theory}. At first glance, the spectra are somewhat blurry below the Fermi level, which means that the Fe-$3d$ electrons are in the incoherent states. This is because we adopted quite high temperature ($\beta = 40.0$) in the DFT + DMFT calculations~\cite{theory}. Once the temperature is lowered, an electronic incoherent-coherent crossover will happen~\cite{Haule_2009}. In Fig.~\ref{fig:akw}(a), a large proportion of electronic excitations are washed out due to the high-temperature effect. But the flat bands are still clearly visualized. Several sets of representative Kagome flat bands and Dirac bands with their orbital characters are plotted in Fig.~\ref{fig:akw}(b)-(e) for a detailed comparison. The bands in the vicinity of the Fermi level obtained by the DFT + DMFT method are strongly renormalized as compared to the corresponding ones obtained by the DFT + SOC method. To be more specific, the bandwidths are reduced, and the Kagome flat bands and Dirac cones are shifted towards the Fermi level obviously. For example, the renormalized flat bands $\alpha$ and $\beta$ locate at approximately -0.10 eV and -0.58 eV, respectively [see Fig.~\ref{fig:akw}(b) and (c)]. If the long-range interactions are ignored, their central energy levels should be around -0.25 eV and -0.69 eV, respectively. Another example is about the Dirac band gaps $\Delta_2$ and $\Delta_3$ [see Fig.~\ref{fig:akw}(d) and (e)]. Their positions are shifted from -1.10 eV and -0.66 eV (without renormalization) to -0.80~eV and -0.50~eV (with renormalization), respectively. Beyond that, since the SOC was not included in the DFT + DMFT calculations~\cite{theory}, only band crossings are seen. Clearly, the band structure is strongly affected by electron correlation. These results also point out a practicable strategy to tune the Kagome bands.     

\begin{figure*}[ht]
\centering
\includegraphics[width=\textwidth]{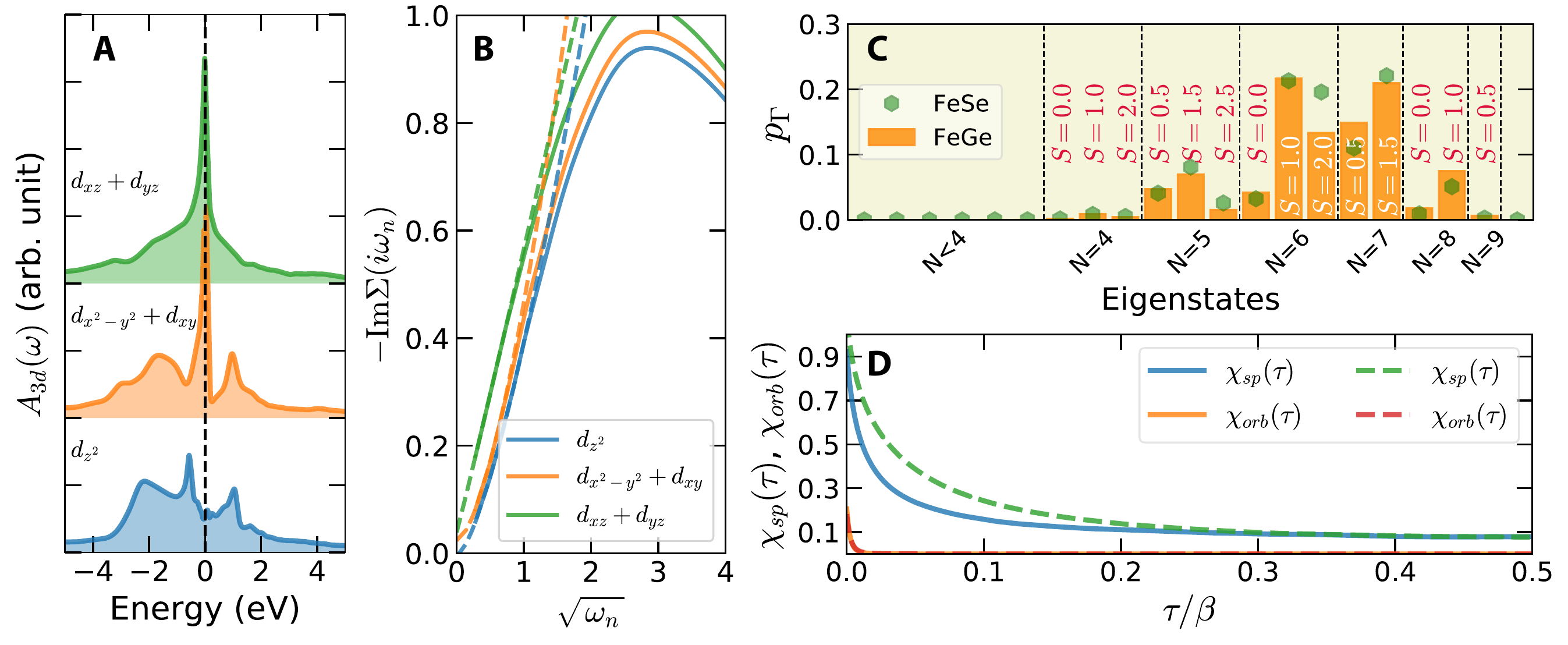}
\caption{(Color online). Hundness in the electronic structure of hexagonal FeGe obtained by DFT + DMFT calculations. (a) Orbital-resolved spectral functions. (b) Imaginary-parts of Matsubara self-energy functions. We adopted Eq.~(\ref{eq:self-energy}) to fit the data at low-energy regime. The fitting results are represented by dashed lines. (c) Valence state histograms. (d) Imaginary-time spin-spin correlation functions $\chi_{sp}(\tau)$ and orbital-orbital correlation functions $\chi_{orb}(\tau)$. The data for FeSe at normal state are shown in this figure as a comparison. They are represented as diamond symbols in panel (c) and dashed lines in panel (d). \label{fig:hund}}
\end{figure*}

%P9: Turn to hundness or mottness. 
We are interested in not only the role play by electron correlation in the electronic structure of hexagonal FeGe, but also its underlying mechanism. In other words, what's the origin of electron correlation in this material? Does it belong to Mottness or Hundness? In order to elucidate this problem, we tried to calculate the spectral function, self-energy function, valence state histogram, spin-spin correlation function, and orbital-orbital correlation function of Fe-$3d$ electrons.     

%P10: Orbital selective electronic correlation.
\emph{Orbital selective electronic correlation.} Fig.~\ref{fig:hund}(a) depicts the orbital-resolved spectral functions for various Fe-3$d$ orbitals. For the $d_{z^2}$ orbital, it shows a pseudogap-like feature at the Fermi level, two shoulder peaks near $\pm$ 0.8~eV, and a broad hump with maxima near -2.1~eV. For the $d_{x^2-y^2} + d_{xy}$ orbitals, a sharp and narrow peak appears at the Fermi level. The hump and side peaks shift their positions slightly towards the Fermi level. For the $d_{xz} + d_{yz}$ orbitals, its spectrum possesses a single incoherent peak at the Fermi level, with effectively overlap with the Hubbard side bands. We find that these spectra are analogous to those observed at Ru's $d_{xy}$ and $d_{xz}+d_{yz}$ orbitals in ruthenates Sr$_{2}$RuO$_{4}$, which is an archetypal strongly correlated material that exhibits Hundness~\cite{deng:2019}.   

%P11: Orbital selective electronic correlation
The orbital-resolved self-energy functions at Matsubara axis are shown in Fig.~\ref{fig:hund}(b). They are all metallic. We adopted the following equation~\cite{PhysRevLett.101.166405} to fit their low-frequency parts:
\begin{equation}
\label{eq:self-energy}
-\text{Im}\Sigma(\omega_n) = C (\omega_n)^{\alpha} + \gamma.
\end{equation}
Finally, we obtained $\alpha \approx$ 0.71, 0.80, and 0.52 for the $d_{z^2}$, $d_{x^2-y^2} + d_{xy}$, and $d_{xz}+d_{yz}$ orbitals, respectively. These values deviate clearly from the prediction of Landau's Fermi-liquid theory, i.e., $\Sigma(i\omega_n) \propto i\omega_n$ at small $\omega_n$~\cite{PhysRevLett.101.166405,RevModPhys.73.797}. Note that $\gamma \equiv -\text{Im}\Sigma(\omega_n \to 0)$, which denotes the low-energy scattering rate. Because the largest $\gamma$ occurs for the $d_{xz}+d_{yz}$ orbitals [see intercepts of the dashed lines at $y$-axis in Fig.~\ref{fig:hund}(b)], so electrons at these orbitals will suffer more effective scatterings. From the self-energy functions at real axis, we can further estimate the renormalization factors $Z$ and electron effective masses $m^{*}$ via the following equation~\cite{RevModPhys.68.13}:
\begin{equation}
Z^{-1} = \frac{m^{*}}{m_e} = 1 - \frac{\partial \text{Re}\Sigma(\omega)}{\partial \omega}\Big|_{\omega = 0}.
\end{equation}
The results are as follows: $Z \approx$ 0.57, 0.49, and 0.30; $m^{*} \approx$ 1.75$m_e$, 2.04$m_e$, and 3.30$m_e$ for the $d_{z^2}$, $d_{x^2-y^2} + d_{xy}$, and $d_{xz}+d_{yz}$ orbitals, respectively. According to these data, we can conclude that the electron correlation in hexagonal FeGe is orbital selective~\cite{Kostin2018,Yin2011}. The out-of-plane $d_{xz}+d_{yz}$ orbitals are more correlated than the others. 

%P12: High-spin states and valence state fluctuations.
\emph{High-spin states and valence state fluctuations.} In strongly correlated materials that exhibit Hundness, high-spin ground states are usually favorite~\cite{Yin2011,STADLER2019365}. Valence state histogram (or atomic configuration probability) is a powerful tool to survey the spin states of correlated electron materials. It represents the probability $p_{\Gamma}$ to find out a given valence electron in any atomic configurations $|\Gamma\rangle$ which are labelled by using some good quantum numbers, such as total occupancy $N$ and total spin $S$ for $d$-electron systems~\cite{PhysRevB.75.155113}. From Fig.~\ref{fig:hund}(c), it is clear that the contributions from those atomic configurations within high-spin states ($S = 1.0$, 1.5, and 2.0) are predominant. They account for approximately 30.1\%, 28.0\%, and 13.8\%, respectively. Thus, the averaged total spin is $\langle S \rangle \approx 1.14$ which is consistent with the previous estimation~\cite{Yin2011}. Another important information we can learn from Fig.~\ref{fig:hund}(c) is the effective occupancy of Fe-3$d$ orbitals. The nominal $3d$ occupancy for Fe$^{2+}$ ions is 6.0. Since strong valence state fluctuations, the effective $3d$ occupancy $\langle N \rangle \equiv \sum_{\Gamma} N_{\Gamma} p_{\Gamma} \approx 6.40$, where $\Gamma$ goes through the atomic configurations with $p_{\Gamma} > 0.0$. 

%P13: Spin-freezing state and spin-orbital separation.
\emph{Spin-freezing state and spin-orbital separation.} We further calculated the imaginary-time spin-spin correlation functions $\chi_{sp}(\tau) \equiv \langle S_z(\tau) S_z(0) \rangle$ and orbital-orbital correlation functions $\chi_{orb}(\tau) \equiv \langle N(\tau) N(0) \rangle $ for Fe-3$d$ electrons. The results are illustrated in Fig.~\ref{fig:hund}(d). In a Fermi-liquid phase at low temperature $T$, $\chi_{sp}(\tau=\beta/2) \sim 1/\beta^2$. However, the calculated $\chi_{sp}(\tau=\beta/2)$ is seen to approach a finite constant value $s$ ($s \gg 1/\beta$), which indicates violation of the Fermi-liquid theory~\cite{RevModPhys.73.797} and emergence of a spin-freezing phase~\cite{PhysRevLett.101.166405}. On the other hand, $\chi_{orb}(\tau)$ is much smaller than $\chi_{sp}(\tau)$. The former approaches zero very quickly at short times $\tau < 1/\beta$. The difference between spin and orbital dynamics disclose an interesting trait that is easily observed at Hund metals: spin-orbital separation~\cite{PhysRevLett.115.136401,STADLER2019365}. In a multi-orbital systems, the spin and orbital degrees of freedom will be screened asynchronously as the temperature is lowered. Generally, orbital screening occurs and completes at much higher temperatures than spin screening, i.e., $T^{onset}_{orb} \gg T^{onset}_{sp} \gg T^{cmp}_{orb} \gg T^{cmp}_{sp}$~\cite{deng:2019}. Thus, in the intermediate temperature regime ($T^{onset}_{orb} > T > T^{cmp}_{sp}$), large spins induced by Hund's rule coupling $J_{\text{H}}$ survive. They fluctuate slowly as if they are frozen, and they are coupled to the screened orbital degrees of freedom. Below $T^{cmp}_{sp}$, both spin and orbital screenings are complete, the Fermi-liquid behavior will be restored. Clearly, hexagonal FeGe falls into the intermediate regime and shows the fingerprints of Hundness. Finally, we would like to remark that the electronic structure of FeGe resembles those of iron-based unconventional superconductors~\cite{RevModPhys.87.855}. For example, tetragonal FeSe has been long recognized as a typical Hund metal~\cite{Yin2011,Haule_2009}. Though its electron correlation is somewhat stronger than that in FeGe and closer to an orbital selective Mott phase~\cite{Kostin2018}, it's Fe-$3d$ electrons still prefer high-spin atomic configurations and exhibits conspicuous spin-orbital separation [see Fig.~\ref{fig:hund}(c)-(d)].      

%% summary
%%%%%%%%%%%%%%%%%%%%%%%%%%%%%%%%%%%%%%%%%%%%%%%%%%%%%%%%%%%%%%%%%%%%%%%%%%

%P14: Summary
In summary, we studied the electronic structure of FeGe in Kagome lattice via the DFT + SOC and DFT + DMFT methods. Our results uncover the coexistence of orbital selective massive Dirac fermions and Kagome-derived flat bands. The electron correlation would play a vital role in the electronic structure of FeGe. First, the Dirac points and flat bands are renormalized and moved towards the Fermi level. Second, the electron correlation in FeGe is strong and orbital-dependent. The out-of-plane $d_{xz}+d_{yz}$ orbitals are more correlated than the $d_{z^2}$ and $d_{x^2-y^2} + d_{xy}$ orbitals. The last and the most important, we identify quite remarkable signatures of Hundness, including the non-Fermi-liquid behavior, high-spin state, spin frozen phase, and spin-orbital separation. All these results suggest that hexagonal FeGe is not only a Kagome metal, but also a Hund metal. It will provide us with a fertile ground to explore the entanglement of topological excitations, Kagome flat bands, and Hund's-coupling-induced correlation.      

\begin{acknowledgments}
We thank Prof. Rui Yu for fruitful discussions. This work was supported by the Natural Science Foundation of China (No.~11874329, 11934020, and 11704347), and the Science Challenge Project of China (No.~TZ2016004).
\end{acknowledgments}

%% reference
%%%%%%%%%%%%%%%%%%%%%%%%%%%%%%%%%%%%%%%%%%%%%%%%%%%%%%%%%%%%%%%%%%%%%%%%%%

\bibliography{hund}

\end{document}